\documentclass[12pt]{article}

\usepackage{graphicx}
\begin{document}

\begin{center}
{\large\bf Regulation of Migration of Chemotactic Tumor Cells by the Spatial Distribution of the Collagen 
Fibers' Orientation}

\bigskip
Youness Azimzade,$^1$ Abbas Ali Saberi,$^{1,\dagger}$ and Muhammad Sahimi$^{2,\ddagger}$

{\it $^1$Department of Physics, University of Tehran, Tehran 14395-547, Iran}

{\it $^2$Mork Family Department of Chemical Engineering and Materials Science, University of Southern 
California, Los Angeles, California 90089-1211, USA}

\end{center}

\bigskip
Collagen fibers, an important component of the extracellular matrix (ECM), can both inhibit and promote 
cellular migration. {\it In-vitro} studies have revealed that the fibers' orientations are crucial to 
cellular invasion, while {\it in-vivo} investigations have led to the development of tumor-associated 
collagen signatures (TACS) as an important prognostic factor. Studying biophysical regulation of cell 
invasion and the effect of the fibers' oritentation not only deepens our understanding of the phenomenon, but
also helps classifying the TACSs precisely, which is currently lacking. We present a stochastic model for 
random/chemotactic migration of cells in fibrous ECM, and study the role of the various factors in it. The 
model provides a framework, for the first time to our knowledge, for quantitative classification of the 
TACSs, and reproduces quantitatively recent experimental data for cell motility. It also indicates that
the spatial distribution of the fibers' orientations and extended correlations between them, hitherto 
ignored, as well as dynamics of cellular motion all contribute to regulation of the cells' invasion length,
which represents a measure of metastatic risk. Although the fibers' orientations trivially affect randomly 
moving cells, their effect on chemotactic cells is completely nontrivial and unexplored, which we study in
this paper. 

\newpage

\begin{center}
{\bf I. INTRODUCTION}
\end{center}

Up to 90 percent of  cancer-associated mortality is attributed to metastasis, but despite this fact, 
metastasis has remained one of the least understood aspects of the disease [1]. During metastasis, cells 
disseminate from the initial tumor, intravasate into the the surrounding vessels, and colonize within a new 
host tissue [2]. Despite development of many prognostic measures for evaluating the metastatic risk, there 
is still intensive ongoing research for gaining deeper understanding of the phenomenon and evaluating 
accurately its risk, in order to minimize treatment failure and costs [3].

Cell migration plays a crucial role in metastasis [4], as the physical translocation during metastasis 
happens through cellular migration that is either directed or random [5]. If the basic machinery of cell 
mobility is activated, but without guiding principle, the cells migrate randomly. In the  presence of 
external/internal guidance cues, however, the cells undergo directed migration [5]. When soluble chemotactic 
agents, such as chemokine and growth factor, represent the external cue, cancer cells may climb the gradient 
and undergo chemotaxis in order to metastasize [6]. As a result, during metastasis, cancer cells migrate 
randomly or are directed until they reach blood vessels and enter its stream. Cellular migration 
{\it in vivo} happens within a heterogeneous environment that is composed mainly of extracellular matrix 
(ECM), which can significantly alter cellular migration [7]. Many models, such as random [8] and persistent 
walks [9], as well as other types of models [10-12] have been developed to describe and/or simulate 
cellular migration.  

The ECM provides the environment that supports cell maintenance [13] and it influences [14] cellular 
migration through its physical properties [14], such as confinement [15-20], fiber topography [21-26], and 
bulk characteristics [27-36]. In particular, orientation of the ECM's fibrils affects cells' direction of 
migration [37-41]. Alignment of the fibrillar matrix, both {\it in vitro} and {\it in vivo} controls 
migration and promotes directional cell migration [42,43], and reorients cell motility without altering its 
overall magnitude [44,45]. Moreover, the fibers can either impede tumor invasion by acting as a barrier 
against migration [46-48], or facilitate it by providing high-speed ``highways'' [49] based on their 
orientation. Experimental investigations have studied the effect of the orientation of the fibers on 
cellular invasion [50-52]. They have indicated larger invasion extent along the direction of aligned ECM's 
fibers for both random and directed migration, but they still fail to provide a quantitative understanding 
of such regulation. Physics-based models based on percolation [53,54] have also been used to simulate the 
ECM structure.

The profound effect of fiber orientation {\it in vivo} has led to the emergence of a prognostic factor, 
known as tumor-associated collagen signature (TACS), which predicts the behavior of tumor based on the 
the type and structure of the ECM alignment [3,55]. According to this approach, there are three types of 
signatures [56]: TACS-1, representing dense wavy collagen fibers (CFs); TACS-2, which is indicative of 
linear CFs parallel to the tumor's border, and TACS-3, identified by the presence of linear CFs 
perpendicular to the tumor's border. Screening of the TACS is a clinical prognostic tool, and TACS-3 could 
indicate poor survival rate, hence suggesting that quantifying CFs alignment may be an independent 
prognostic marker [3,55,57]. But, despite the significance of the classification as a strong prognostic 
factor, as well as a quantitative approach to study alignment of the ECM fibers [58], it remains 
qualitative [55] because, for example, it is not clear what angle between the fibers and the tumors' border
constitutes the "dividing angle" between the TACS-2 and TACS-3, and how the transition between the two 
occurs. Thus, fiber classification should be addressed by development of quantitative understanding of the 
effect of the orientations of the fibers on cell migration. In this paper we describe a new model, and 
utilize analytical arguments and numerical simulation to study the effect of the ECM's fibers' structure 
on cell migration, which provide quantitative understating of the ECM's fibers dividing angle.  

The rest of this paper is organized as follows. In Sec. II we describe the details of the model that we use 
in our study. The results are presented and discussed in Sec. III, while the paper is summarized in Sec. IV.
 
\begin{center}
{\bf II. THE MODEL}
\end{center}

Physical cues of the cellular environment, such as organization of the ECM components, affect the cells and 
their motility, and force transduction [59]. Collagen alignment regulates migration by directing cellular 
protrusions along aligned fibers [60]. The alignment also promotes directed migration by a combination of 
traction forces and contact guidance mechanisms [61]. Fibrillar topographical cues in the form of 
one-dimensional (1D) nanofibres guide cell migration {\it in vivo} [62]. As such, the cells migrate 
directionally along oriented fibers [21]. The size of the CFs and the structure of the ECM exhibit wide 
diversity, however. Experimental studies [44,45] have attempted to mimic the observed structures [50-52], 
and have indicated that the fibers have an average length of 20 $\mu$m, with their orientations following a 
Gaussian distribution. In this paper we rely on such data as the basis of our model. 

\begin{figure}[!htb]
\centerline{\includegraphics[width=0.95\linewidth]{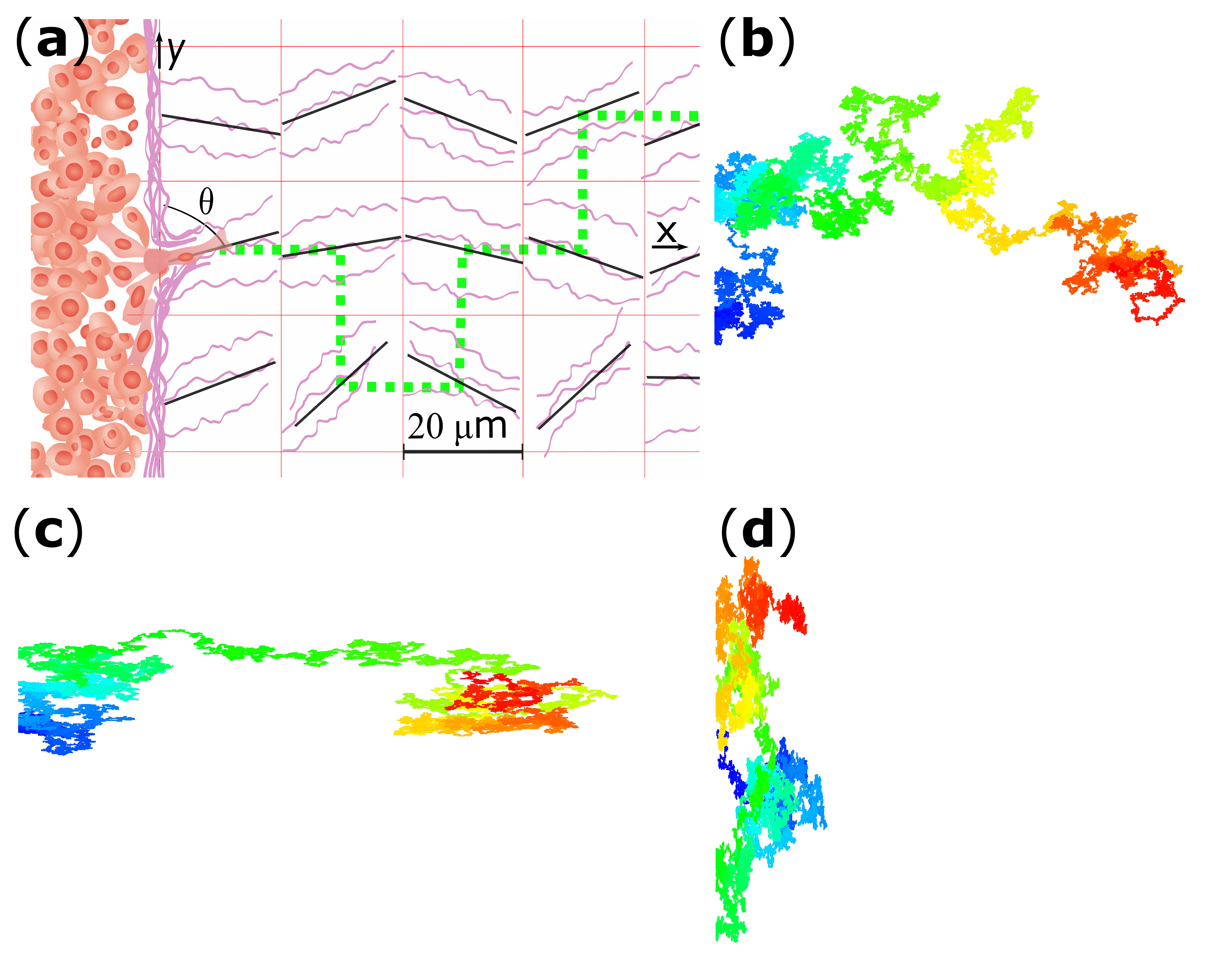}}
\caption{(a) A cell and the medium into which it migrates from a torn boundary at (0,0) (middle left). The 
fibers are shown by the solid lines. Once one step of migration takes place, the cell will be in one of the
nearest neighbor units and continues to migrate according to the same probabilities. The green dashed 
line shows a possible random trajectory for the migration. Sample trajectories are for ($\sigma,\;
\bar{\theta}$) (b) ($\pi/2,\;\pi/2$); (c) ($\pi/10,\;\pi/2$), and (d) ($pi/10,\;0$).}
\end{figure}

To model the ECM's structure, we divide the cellular environment into $20\times 20\;\mu$m$^2$ units,
corresponding to the length of a fiber, and assign a direction to each unit along the fiber inside it; see 
Fig. 1(a). In principle, the size of the fibers can be different for different tissues, but the qualitative
aspects of the results would be the same, if we use different fiber size for different tissues. It would also
not be difficult to use different fiber size for different tissues. Note also that the exact length of the 
fibers does not play an important direct role in the simulations. What is significant is the existence
of the fibers in the ECM that provide a medium for the cells to advance.

The ECM is, of course, three dimensional. There is also extensive evidence that extrapolation from 2D is far 
from straightforward [5]. But, as emphasized by others [21], the fact is that the fibers have a 1D structure.
As such, we argue that migration happens on 1D structures embedded in a higher-dimensional space. Given this
assumption, then, regardless of dimensions of the space in which the fibers are embedded, our results should 
be valid. In fact, in the experiments with which we compare our results (see below), the cellular medium 
is quasi-2D. Migration happens on the 1D collagen fibers, and the height (thickness) of the medium is so
small compared to the other dimensions that the cellular medium is essentially 2D. If we extend the cellular 
environment in our model in the third direction $z$ by keeping the same configuration of the fibers around 
the $x$ axis (shown in Fig. 1), the result for the motility would remain the same. 

The heterogeneous cellular environment is not completely random, but contains extended correlations [63] 
the existence of which has been confirmed by {\it in-vivo} studies [55], although the structure of the 
correlation function has not been characterized yet. To generate a distribution of the fibers with spatial 
correlations, we use the fractional Brownian motion (FBM) [64] according to which for the fiber orientations 
at {\bf x} and {\bf x'} one has, $\langle(\theta_{\bf x}-\theta_{{\bf x}'})^2\rangle\propto|{\bf x}-
{\bf x}'|^{2H}$. Here, $0<H<1$ is the Hurst exponent [65] with $H>0.5$ representing positive correlations so 
that the fibers' alignments vary smoothly as $H\to 1$, whereas negative correlations are represented by 
$H<0.5$ and, thus, the orientations fluctuate widely as $H\to 0$. Note that we do not claim that the FBM 
represents the actual type of the correlations, rather we use it as a typical stochastic functions that 
produces extended correlations. At the same time, the FBM has found many applications in biological 
phenomena [66-70]. Note that the model that we use is not for migration in 2D substrates. Instead, it
imitates migration on collagen fibers, which for convenience is considered to be two dimensional. In a 3D 
model one would have $20\times 20\times 20$ $\mu$m$^3$ cubes, in which we assign $\theta$ as the angle 
with the normal to the tumor's border plain, and a second angle $\phi$ for the orientation in the tumor's 
border plain.

As Fig. 1(a) indicates, the tumor's boundary is at ${\bf x}=0$. Then, if a cell is in a unit with a fiber 
orientation $\theta$ with respect to the $y$ axis [vertical axis in Fig. 1(a)], the probabilities of 
migrating in the $x$ and $y$ directions are, respectively, $\sin^2\theta$ and $\cos^2\theta$. In the 
corresponding 3D model we set the probabilities to be $\sin^2\theta$, $\cos^2\theta\cos^2\phi$, and $\cos^2
\theta\sin^2\phi$ for $x$, $y$ and $z$ directions, respectively, where $\phi$ desribes the direction in the 
plane parallel to tumor's border. The probabilities provide us with a means of understanding the effect on 
cell migration of the fibers' alignment and their distribution. Note that the orientations lead to cellular 
alignment that causes directed migration [44,45]. We assume that the migration probabilities capture the 
effects of both the ECM's and the cellular alignment. Then, the orientations are selected according to the 
FBM with a standard deviation of $\sigma$ around $\bar{\theta}$.

To study chemotaxis we use the normalized barrier [71] or the Keller-Segel [72] model, according to which 
for a cell moving in a 1D medium in a constant external chemical potential gradient, the probabilities of 
moving right and left are $r=p$ and $l=1-p$, where $|p-1/2|$ ($|p-1/4|$ in 2D) is the strength of 
chemotaxis that is regulated by the gradient strength and the cells' ability to detect and respond to it. 
Extending the model to 2D with no chemtaxis in the $y$ direction leads to $r=p$ and $l=u=d=(1-p)/3$, in 
which $u$ and $d$ are the probabilities of moving up and down. Coupling between the effect of chemotaxis 
and the ECM alignment is implemented by considering $r\propto p\sin^2\theta$, $l\propto (1/3)(1-p)\sin^2
\theta$, and $u=d\propto (1/3)(1-p)\cos^2\theta$, which after normalization lead to, 
\begin{eqnarray}
& & r=\frac{3p\sin^2\theta}{S}\;,\\
& & l=\frac{\sin^2\theta(1-p)}{S}\;,\\ 
& & u=d=\frac{\cos^2\theta(1-p)}{S}\;,
\end{eqnarray}
with the drift velocity $v_x$ given by, 
\begin{equation}
v_x=\frac{\delta(r-l)}{\tau}=\frac{\delta\sin^2\theta(4p-1)}{D}\;,
\end{equation}
and $v_y=0$, where $S=2(1-p)\cos^2\theta+(1+2p)\sin^2\theta$ and $D=S\tau$, with $\delta=20\;\mu$m being 
the jump's length (the units' size), and $\tau$ is the duration of a single jump. Though we do not present 
results in which time is explicitly present, consistent with the experiments [44,45] on cells' velocity, we 
set $\tau=1$ hour in the simulations, and consider the drift only in the positive $x$ direction, whereas in 
the $y$ direction only diffusion with no drift occurs.

\begin{center}
{\bf III. RESULTS AND DISCUSSIONS}
\end{center}

We first consider the cells to be initially at ${\bf x}=0$, the tumor boundary, moving for a large number of 
time steps in the cellular medium with the FBM disribution of the fibers' orientations with given $\sigma$ 
and $\bar{\theta}$. At each time step a cell moves according to the probabilities given by Eqs. (1) - (3). 
It is free to move anywhere, except crossing the ${\bf x}=0$ line. Examples of the trajectories of the cells 
are presented in Figs. 1(b)-1(d).

We then check if the model reproduces recent experimental data for cell motility in various directions. 
Defining the directional motilities by $\mu_x=\langle x^2\rangle/t$ and $\mu_y=\langle y^2\rangle/t$, recent 
experimental studies on the ECM's fiber alignment with $\bar{\theta}=\pi/2$ and $\sigma=0.13\pi$ reported 
that [44,45], $\mu_x/\mu_y\approx 5$. Figure 2(a) presents the simulation results with the same parameters. 
We find that after long enough times, $\langle x^2\rangle\approx 420$, $\langle y^2\rangle\approx 90$ and, 
therefore, $\mu_x/\mu_y=\langle x^2\rangle/\langle y^2\rangle=420/90\approx 4.67$, in excellent agreement 
with the experimental data.

Since our model is 2D, but the experiments had been carried out in seemingly 3D media, the agreement may 
seem to be fortuitous. Note, however, that the experiments were actually carried out in a 
quasi-two-dimensional cellular medium, and although migration occurs on the collagen fibers embedeed in 3D
space, the evironment is limited in third direction $z$. But, even if the cellular medium were truly 3D, 
$\langle x^2\rangle$ should remain unchanged, because in the 3D model we define the motility in the border 
plain by $\mu_r=\mu_y+\mu_z$. In that case we would still have $\mu_x/\mu_r\approx 5$, hence indicating that 
our result, at least for the uncorrelated cellular media, would not change if we use a fully 3D model.

\begin{figure} [!htb]
\centerline{\includegraphics[width=0.8\linewidth]{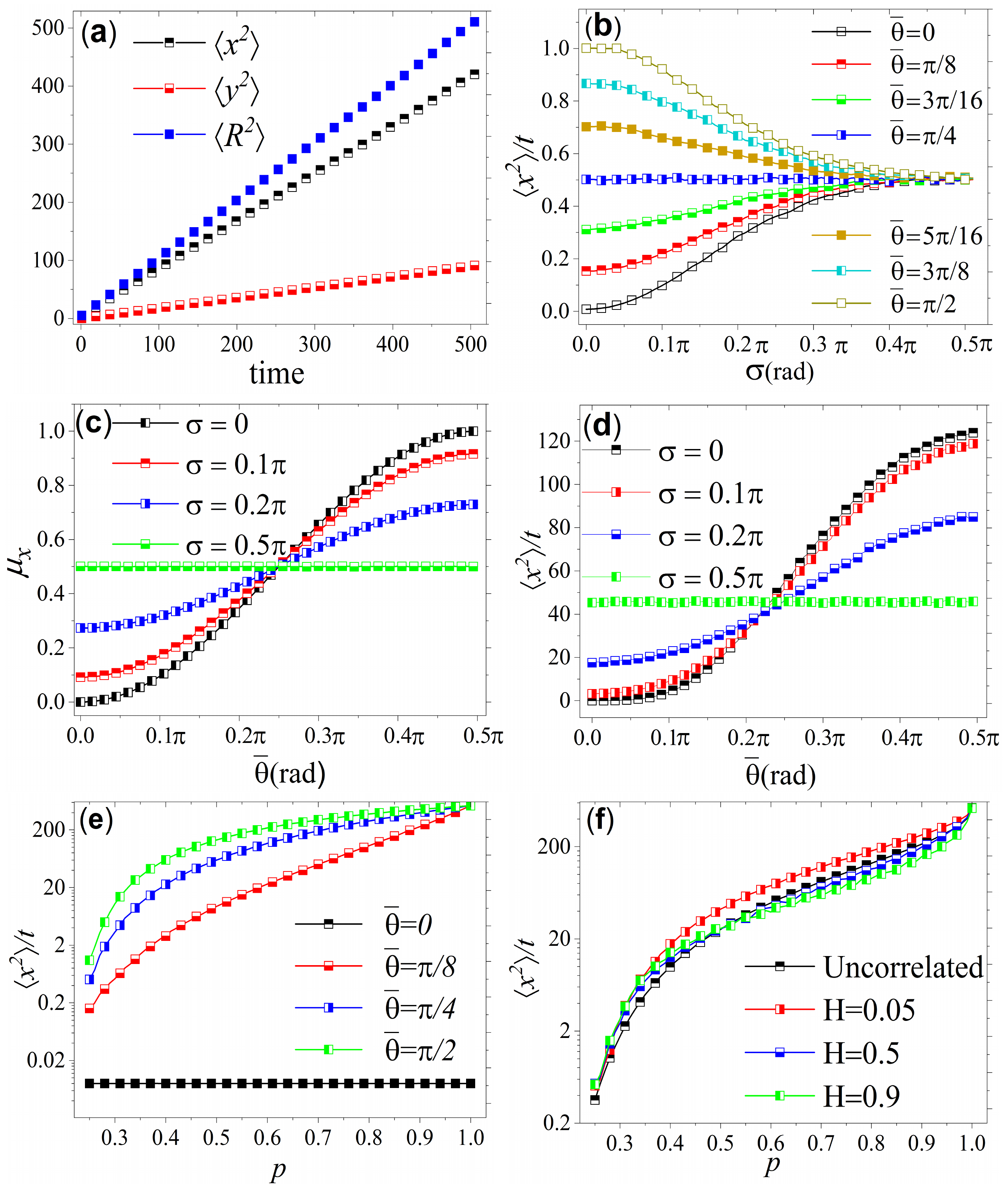}}
\caption{(a) Evolution of $\langle x^2\rangle$, $\langle y^2\rangle$, and $\langle R^2\rangle=\langle x^2
\rangle+\langle y^2\rangle$. (b) Effect of $\sigma$ of the orientation distribution. (c) Dependence of the 
motility $\mu_x$ on $\sigma$ and $\bar{\theta}$ for $p=1/4$, defining the crossing point $\theta^*$. (d) 
Same as (c) but with $p=0.8$. (e) Effect of directionality, represented by the probability $p$. (f) Same 
as in (e) but in a correlated cellular medium with $\sigma=0.1$ and $\bar{\theta}=\pi/8$.}
\end{figure}

More generally, however, it is important to understand how the spatial distribution of the fibers'
orientations affect the extent of cell invasion. Since migration perpendicular to the tumor boundary at 
${\bf x}=0$ plays the main role in metastasis, we take the net mean displacement, $\langle x^2\rangle^{1/2}$,
as the extent or length of the invasion, which is indicative of the metastatic risk. Consider, first, the 
non-chemotactic case with $p=1/4$. The simulations indicate that as $\sigma$, the standard deviation of the 
orientations' distribution, increases in an uncorrelated medium, $\langle x^2\rangle/t$ converges to the 
same value of about 0.5 for all $\bar{\theta}$, which is expected for a random walk in a homogeneous medium; 
see Fig. 2(b). Precisely the same behavior develops for the chemotactic case, $p>1/4$.

In a cellular environment with well-aligned fibers and small $\sigma$, the dependence on $\bar{\theta}$ of 
all $\langle x^2\rangle/t$ exhibits sigmoidal behavior with a common crossing point at a specific orientation
$\theta^*$, defined by $\langle x^2(\theta^*)\rangle=\langle x^2(\pi/2)\rangle/2$. Note that although the 
shape of $\langle x^2(\bar{\theta})\rangle$ may vary in various limits, it still is an important property 
for quantitative understanding of the effect of the fibers' alignment. Figure 2(c) shows that in random cell 
advancement the transition is at $\bar{\theta}^*=\pi/4$, which provides the first criterion for separating 
TACS-2 and TACS-3: for randomly-advancing tumor cells, fibers with orientations $\theta<\pi/4$ are classified
as TACS-2, and those with $\theta>\pi/4$ as TACS-3. Other definitions of $\theta^*$ may be considered, but 
our approach allows us to classify the fibers based on any reasonable definition of $\theta^*$. Note that
we do not claim, at this point, that the value of $\theta^*$ is universal, independent of the type of fibers'
oritentation distribution, or the dimensionality of the cellular medium. The important point that we would 
like to emphasize is rather the {\it existence} of such a critical angle. We will return to this point 
shortly.  

Consider, next, chemotactic migration with $p>1/4$ for which $\langle x^2\rangle\ne\mu_xt$. Thus, we focus 
on $\langle x^2\rangle/t$, which is a measure of mobility (``diffusivity'') of the cells, and study its 
dependence on $\sigma$ and $\bar{\theta}$. We first note that chemotactic migration does have a transition 
point $\theta^*$; see Fig. 2(d) that presents the results for $p=0.8$. But, before analyzing the
characteristics of $\theta^*$, let us consider the effect of the directionality of cell migration on 
$\langle x^2\rangle$, which is regulated by $p$. Using the expressions for the probabiliies of motion and
the drift velocities, we obtain, $\langle x^2\rangle=\mu_x t+v_x^2t^2=2\langle\sin^2\theta\rangle(1-p)t/3
+[t\langle\delta\sin^2\theta(4p-1)/D\rangle]^2$.

\begin{figure} [!htb]
\centerline{\includegraphics[width=0.85\linewidth]{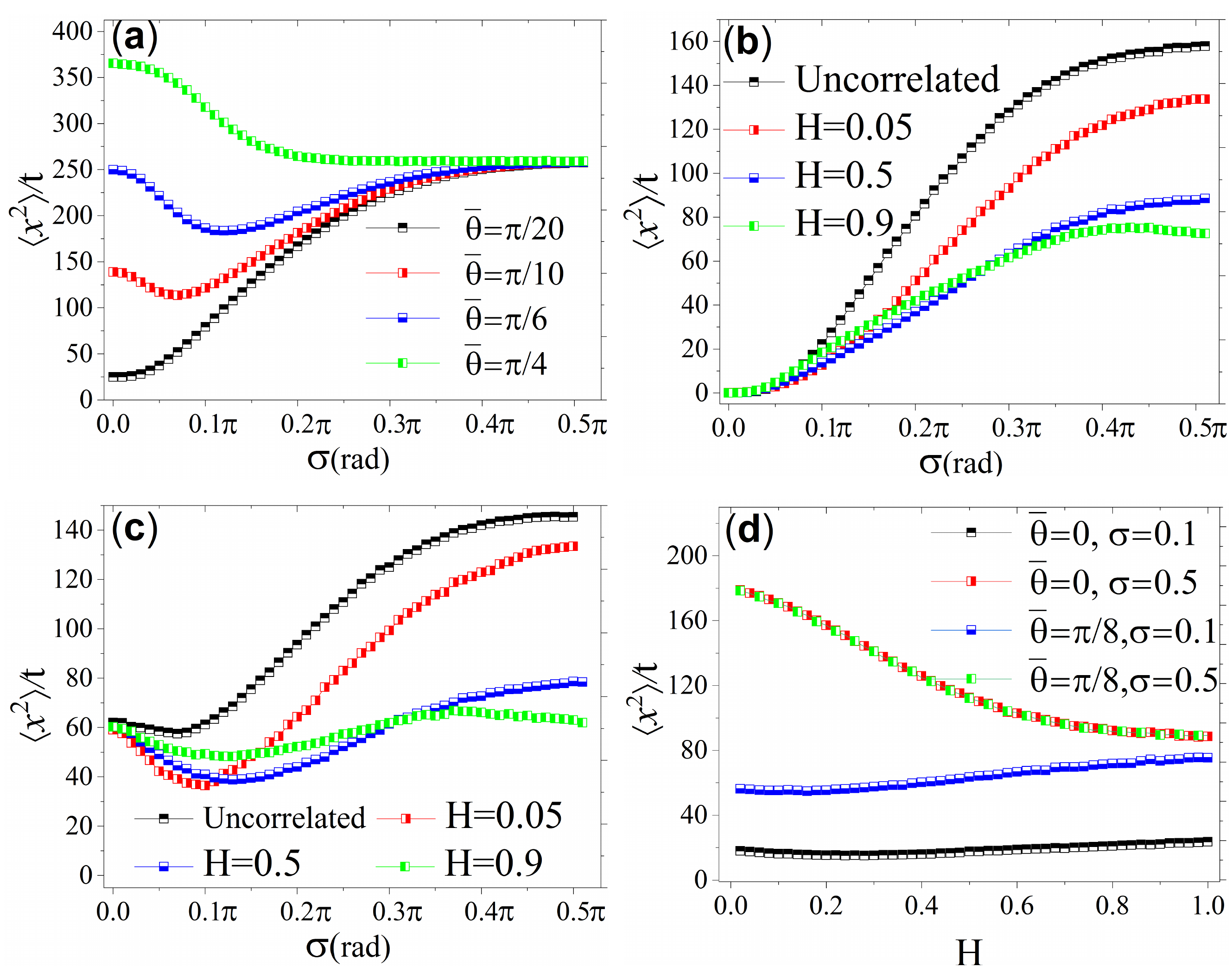}}
\caption{Dependence of $\langle x^2\rangle/t$ on (a) $\sigma$ and $\bar{\theta}$ for $p=0.9$. Increasing 
$\sigma$ decreases the differences between the results for various $\bar{\theta}$; (b) $\sigma$ and the 
Hurst exponent $H$ for $\bar{\theta}=0$, showing that cells move slower in environments with positive 
correlations ($H>0.5$); (c) $\sigma$ and $H$ for $\bar{\theta}=\pi/8$ and $p=0.75$. The correlations affect
the mobility, but the nonmonotonic behavior of $\langle x^2\rangle/t$ remains unchanged, and (d) $H$ for 
$p=0.8$.}
\end{figure}

To understand the effect of $p$ we first considered uncorrelated cellular media and varied both $p$ and 
$\bar{\theta}$. For a given $\bar{\theta}$ $\langle x^2\rangle/t$ increases with increasing $p$; see Fig. 
2(e). Depending on $p$, the correlations may increase or decrease $\langle x^2\rangle/t$; see Fig. 2(f). 
Thus, the correlations are indeed relevant and regulate $\langle x^2\rangle$. Moreover, regardless of the 
correlations' type ($H<0.5$ or $H>0.5$), directionality of the cell advancement does influence $\langle x^2
\rangle$ and $\theta^*$ significantly.

To study the effect of $\sigma$, we computed $\langle x^2\rangle$ for $p=0.9$ and several values of 
$\bar{\theta}$. As Fig. 3(a) shows, for large $\sigma$ all $\langle x^2\rangle/t$ converge to the same 
eventual value. But, more interestingly and contrary to the limit $p=1/4$, the dependence of $\langle x^2
\rangle/t$ on $\sigma$ is not trivial. It initially decreases and then increases. Such non-monotonic 
behavior may be understood by noting that the average probability $\langle r\rangle$ of moving in the 
positive $x$ direction does not vary monotonically with $\sigma$. Figure 3(a) also indicates that, although
all the $\langle x^2\rangle/t$ eventually converge to the same value, the shape of their variations depends
on the value of $\sigma$.

\begin{figure} [!htb]
\centerline{\includegraphics[width=0.84\linewidth]{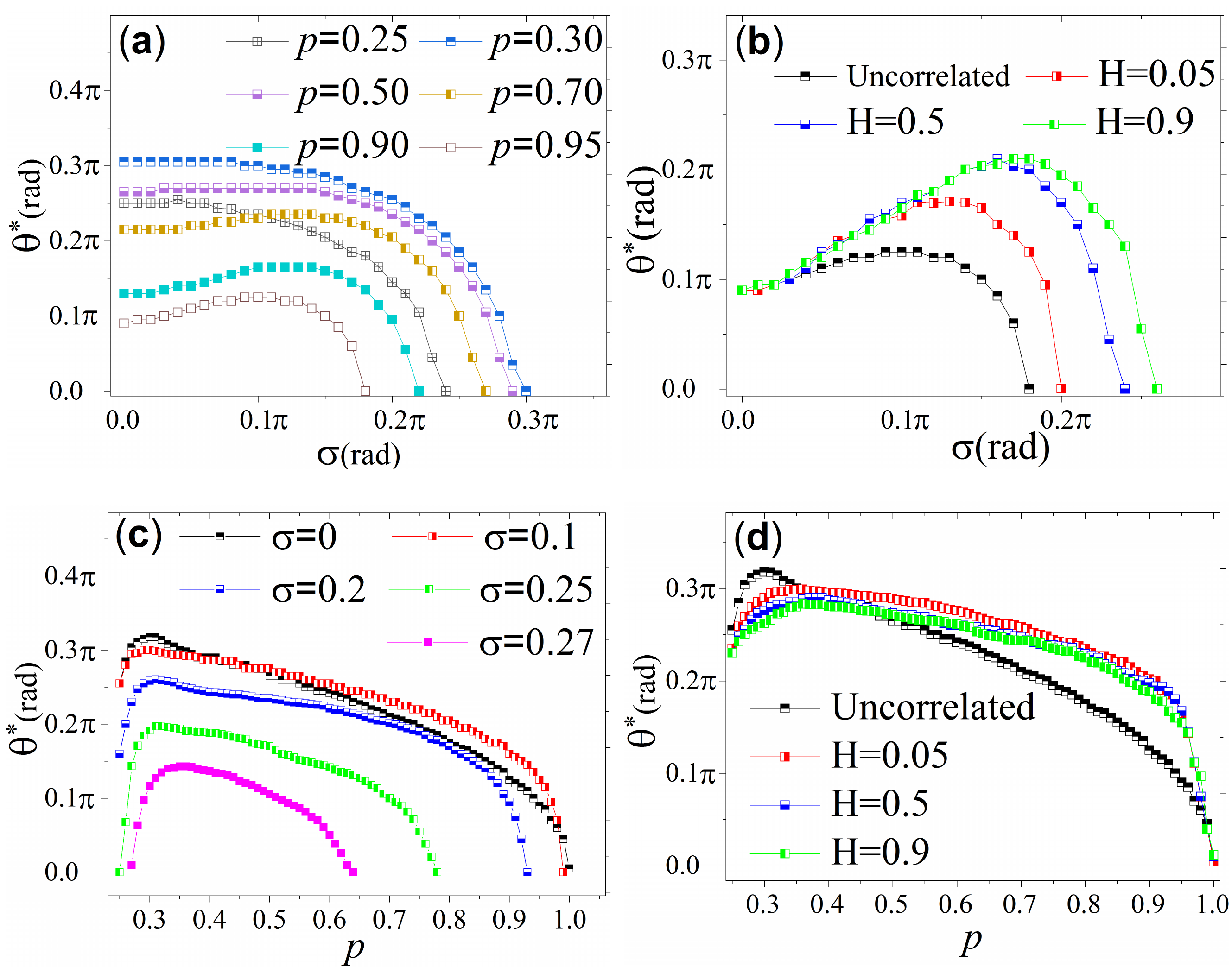}}
\caption{Dependence of $\theta^*$ on (a) $\sigma$ and $p$, showing that varying $p$ slightly changes the 
magnitude of $\theta^*$ and its dependence on $\sigma$, but the main qualitative behavior remains almost the same 
for all $p$; (b) $\sigma$ and $H$ for $p=0.95$; (c) $p$ and $\sigma$, indicting that directionality of the 
motion may increase or decrease $\theta^*$, and (d) $p$ for various $H$.}
\end{figure}

One goal of this paper is to understand the effect of the orientations' correlations on the results, and as 
Fig. 3(b), which presents the dependence of $\langle x^2\rangle/t$ on $\sigma$ for various Hurst exponents 
$H$ in the limits $\bar{\theta}=0$ and $p=0.9$, indicates, the effect is completely nontrivial. As Fig. 3(b) 
indicates, Not only is the growth of $\langle x^2\rangle/t$ with $\sigma$ completely different from those in 
Fig. 3(a), it also indicates that the cells advance more slowly in a cellular environment with positive 
correlations ($H>0.5$), which should be compared with Fig. 3(c) for $p=0.75$ and $\bar{\theta}=\pi/8$. In 
this case, the correlations give rise once again to nonmonotonic dependence of $\langle x^2\rangle/t$ on 
$\sigma$, hence playing a major role in regulating $\langle x^2\rangle$. As Fig. 3(d) indicates, the 
correlations may increase or decrease $\langle x^2\rangle/t$, depending on $\sigma$ and $\bar{\theta}$.

As discussed earlier, the limits $p=1/4$ and $\theta^*=\pi/4$ represent a crossing angle, a sort of a
transition point. Thus, the characteristics of $\theta^*$ as the dividing angle at which a transition from 
low-risk (TACS2) to high-risk (TACS3) metastasis occurs are important. As Fig. 4(a) shows, for every $p$ in 
a noncorrelated medium $\theta^*$ vanishes with increaing $\sigma$, hence indicating that in a cellular
medium with a rather wide $\sigma$, the risk of metastasis increases significantly for almost all the fiber 
directions and migration modes. This provides a framework for classification of the TACSs. Equally 
importantly, Fig. 4(b) indicates that the correlations have a nontrivial effect on $\theta^*$. Figures 4(a)
and 4(b) both indicate that the transition orientation $\theta^*$ is not universal and does depend on the
details of the oritentation distribution and other parameters.

\begin{figure} [!htb]
\centerline{\includegraphics[width=0.9\linewidth]{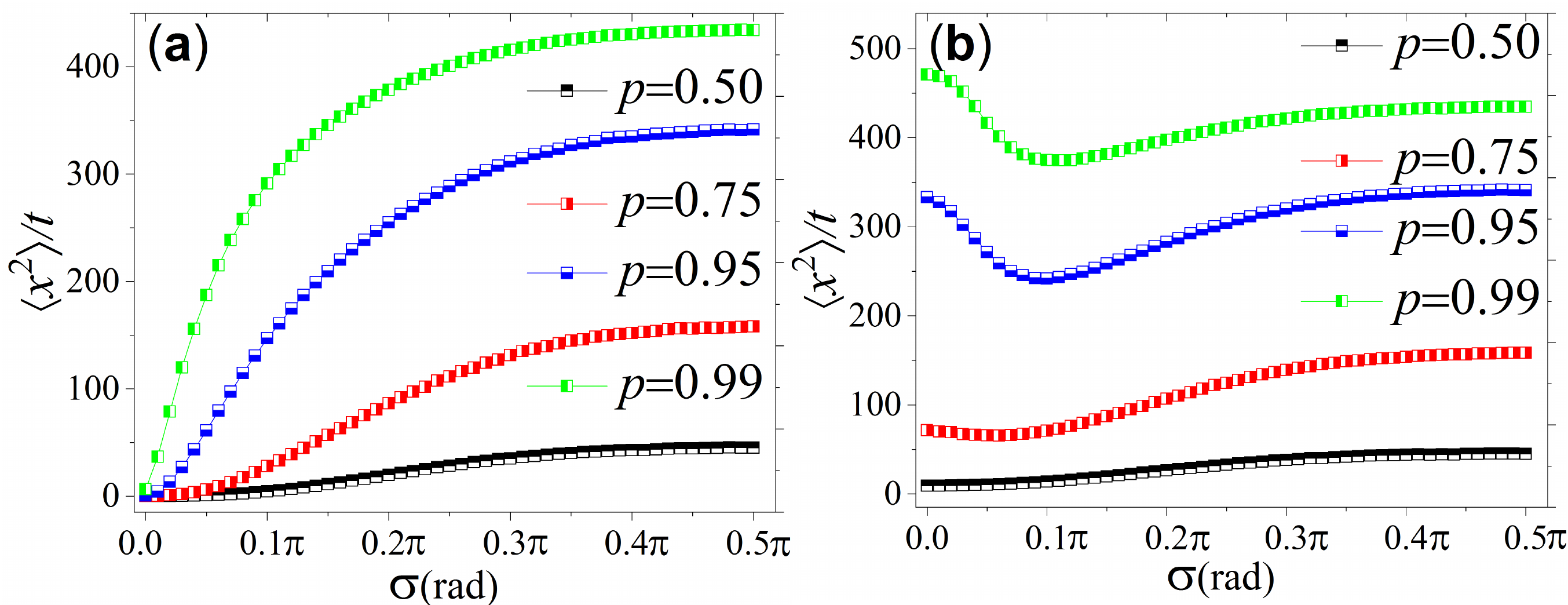}}  
\caption{Dependence of $\langle x^2\rangle/t$ on $\sigma$ and $p$ for (a) $\bar{\theta}=0$, and (b) 
$\bar{\theta}=\pi/8$.}
\label{FIG5}
\end{figure}

Since cells may migrate in various modes, the question of how they alter $\theta^*$ is also important. As 
Fig. 4(c) indicates, directionality of the tumor advancement initially increases and then decreases 
$\theta^*$. The significance of these results is in demonstrating that, while for random motion of the cells 
the fibers' alignment is the main contributing factor to the magnitude of the invasion length, one needs 
a more precise and better defined framework for chemotaxtic migration, as the physical parameters that 
affect the phenomenon, such as the extent and type of the correlations between the fibers' orientations and
the migration mode, play major roles.

We also studied the effect of the probability $p$ on the mobility $\langle x^2\rangle/t$. Figures 5(a) and 
5(b) present, respectively, the results for $\bar{\theta}=0$ and $\pi/8$. Thus, even a small change in the 
mean angle of the fibers' orientations gives rise to remarkable changes in the mobility. In particular, 
decreasing the probability $p$ of moving forward for $\bar{\theta}=\pi/8$ gives rise eventually to a 
nonmonotonic variations of $\mu_x$ with $\sigma$.

\begin{center}
{\bf IV. SUMMARY AND CONCLUSIONS}
\end{center}

We described what we believe is the first model that introduces a dividing angle for classification of 
tumor-associated collagen signature (TACS), demonstrating how various physical features, such as the spatial 
distribution of the extre-cellumar matrix's fibers' orientations and the cells' migration dynamics regulate 
the cell invasion length and, therefore, the metastatic risk. The distribution of the orientations of the 
fibers plays a crucial role, and may promote or inhibit cell migration. The cells' migration mode, ranging 
from random walks to entirely biased walks, also affects the invasion length. Thus, the three factors should 
be considered together for complete classification of the TACSs. This may explain why classification of tumor
environment based solely on the fibers' alignment has not proven to be fruitful for all the cases. 

We emphasize that cell migration is a complex process regulated by biochemical communications between the 
cells and various constituents of the host tissue, as well as the biophysical interactions. Our goal in this 
paper was to investigate the effect of physical interactions. Gaining a comprehensive understanding of the 
ECM regulation of cell migration and the metastasis should integrate all the chemical and physical aspects.

\bigskip

\noindent $^\dagger$ab.saberi@ut.ac.ir

\noindent $^\ddagger$moe@usc.edu

\newcounter{bean}
\begin{list}%
{[\arabic{bean}]}{\usecounter{bean}\setlength{\rightmargin}{\leftmargin}}

\item C.L. Chaer and R.A. Weinberg, A perspective on cancer cell metastasis, Science {\bf 331}, 1559 (2011).

\item C.A. Klein, The metastasis cascade, Science {\bf 321}, 1785 (2008).

\item A. Case {\it et al.}, Identification of prognostic collagen signatures and potential therapeutic 
stromal targets in canine mammary gland carcinoma, PLOS one {\bf 12}, e0180448 (2017).

\item F. van Zijl, G. Krupitza, and W. Mikulits, Initial steps of metastasis: cell invasion and endothelial 
transmigration, Mutat. Res. {\bf 728}, 23 (2011).

\item R.J. Petrie, A.D. Doyle, and K.M. Yamada, Random versus directionally persistent cell migration, Nat. 
Rev. Mol. Cell Biol. {\bf 10}, 538 (2009).

\item E.T. Roussos, J.S. Condeelis, and A. Patsialou, Chemotaxis in cancer, Nat. Rev. Cancer {\bf 11}, 573 
(2011).

\item W.J. Polacheck, I.K. Zervantonakis, and R.D. Kamm, Tumor cell migration in complex microenvironments,
Cellular Mol. Life Sci. {\bf 70}, 1335 (2013).

\item P.-H. Wu, A. Giri, S.X. Sun, and D. Wirtz, Three-dimensional cell migration does not follow a random 
walk, Proc. Natl. Acad. Sci. USA {\bf 111}, 3949 (2014).

\item P.-H. Wu, A. Giri, and D. Wirtz, Statistical analysis of cell migration in 3D using the anisotropic 
persistent random walk model. Nat. Protoc. {\bf 10}, 517 (2015).

\item S. Evje, An integrative multiphase model for cancer cell migration under influence of physical cues 
from the microenvironment, Chem. Eng. Sci. {\bf 165}, 240 (2017).

\item C. Deroulers, M. Aubert, M. Badoual, and B. Grammaticos, Modeling tumor cell migration: From 
microscopic to macroscopic models, Phys. Rev. E {\bf 79}, 031917 (2009).

\item A.J. Loosley, X.M. O'Brien, J.S. Reichner, and J.X. Tang, Describing directional cell migration with a 
characteristic directionality time, PLOS One {\bf 10}, e0127425 (2015).

\item C. Frantz, K.M. Stewart, and V.M. Weaver, The extracellular matrix at a glance, J. Cell Sci. {\bf 123},
4195 (2010).

\item G. Charras and E. Sahai, Physical influences of the extracellular environment on cell migration, Nat. 
Rev. Mol. Cell Biol. {\bf 15}, 813 (2014).

\item K. Wolf {\it et al.}, Physical limits of cell migration: control by ECM space and nuclear deformation 
and tuning by proteolysis and traction force, J. Cell Biol. {\bf 201}, 1069 (2013).

\item Y.-J. Liu, M. Le Berre, F. Lautenschlaeger, P. Maiuri, A. Callan-Jones, M. Heuz\'e, T. Takaki, R. 
Voituriez, and M. Piel, Confinement and low adhesion induce fast amoeboid migration of slow mesenchymal 
cells, Cell {\bf 160}, 659 (2015).

\item A. Rahman-Zaman, S. Shan, and C.A. Reinhart-King, Cell migration in microfabricated 3D collagen 
microtracks is mediated through the prometastatic protein girdin, Cell. Mol. Bioeng. {\bf 11}, 1 (2018).

\item A. Rahman, S.P. Carey, C.M. Kraning-Rush, Z.E. Goldblatt, F. Bordeleau, M. C. Lampi, D. Y. Lin, 
A.J. Garc\'ia, and C.A. Reinhart-King, Vinculin regulates directionality and cell polarity in two- and 
three-dimensional matrix and three-dimensional microtrack migration, Mol. Biol. Cell {\bf 27}, 1431 (2016).

\item S.P. Carey, A. Rahman, C.M. Kraning-Rush, B. Romero, S. Somasegar, O.M. Torre, R.M. Williams, and 
C.A. Reinhart-King, Comparative mechanisms of cancer cell migration through 3D matrix and physiological 
microtracks, Am. J. Physiol. Cell Physiol. {\bf 308}, C436 (2015).

\item C.M. Kraning-Rush, S.P. Carey, M.C. Lampi, and C.A. Reinhart-King, Microfabricated collagen tracks 
facilitate single cell metastatic invasion in 3D, Integr. Biol. (Camb.) {\bf 5}, 606 (2013).

\item A.D. Doyle, F.W. Wang, K. Matsumoto, and K.M. Yamada, One-dimensional topography underlies 
three-dimensional fibrillar cell migration, J. Cell Biol. {\bf 184}, 481 (2009).

\item C.M. Kraning-Rush and C.A. Reinhart-King, Controlling matrix stiffness and topography for the study of 
tumor cell migration, Cell Adh. Migr. {\bf 6}, 274 (2012).

\item L.A. Hapach, J.A. VanderBurgh, J.P. Miller, and C.A. Reinhart-King, Manipulation of {\it in vitro} 
collagen matrix architecture for scaffolds of improved physiological relevance, Phys. Biol. {\bf 12}, 061002 
(2015).

\item M.C. Lampi, M. Guvendiren, J.A. Burdick, and C.A. Reinhart-King, Photopatterned hydrogels to 
investigate the endothelial cell response to matrix stiffness heterogeneity, ACS Biomater. Sci. Eng. {\bf 3},
3007 (2017).

\item F. Bordeleau, L.N. Tang, and C.A. Reinhart-King, Topographical guidance of 3D tumor cell migration at 
an interface of collagen densities, Phys. Biol. {\bf 10}, 065004 (2013).

\item S. Rhee, J.L. Puetzer, B.N. Mason, C.A. Reinhart-King, and L.J. Bonassar, 3D bioprinting of spatially 
heterogeneous collagen constructs for cartilage tissue engineering, ACS Biomater. Sci. Eng. {\bf 2}, 1800 
(2016).

\item K.R. Levental {\it et al.}, Matrix crosslinking forces tumor progression by enhancing integrin 
signaling, Cell {\bf 139}, 891 (2009).

\item M.J. Paszek {\it et al.}, Tensional homeostasis and the malignant phenotype, Cancer Cell {\bf 8}, 241 
(2005).

\item T.A. Ulrich, E.M. de Juan Pardo, and S. Kumar, The mechanical rigidity of the extracellular matrix 
regulates the structure, motility, and proliferation of glioma cells, Cancer Res. {\bf 69}, 4167 (2009).

\item A. Pathak and S. Kumar, Independent regulation of tumor cell migration by matrix stiffness and 
confinement, Proc. Natl. Acad. Sci. USA {\bf 109}, 10334 (2012).

\item F. Bordeleau {\it et al.}, Matrix stiffening promotes a tumor vasculature phenotype, Proc. Natl. Acad.
Sci. USA {\bf 114}, 492 (2017).

\item M.C. Lampi, C.J. Faber, J. Huynh, F. Bordeleau, M. R. Zanotelli, and C.A. Reinhart-King,  Simvastatin 
ameliorates matrix stiffness-mediated endothelial monolayer disruption, PLOS One {\bf 11}, e0147033 (2016).

\item S. Lin, L.A. Hapach, C. Reinhart-King, and L. Gu, Towards tuning the mechanical properties of 
three-dimensional collagen scaffolds using a coupled fiber-matrix model, Materials {\bf 8}, 5376 (2015).

\item F. Bordeleau, J.P. Califano, Y.L.N. Abril, B.N. Mason, D.J. LaValley, S.J. Shin, R.S. Weiss, and 
C.A. Reinhart-King, Tissue stiffness regulates serine/arginine-rich protein-mediated splicing of the extra 
domain B-fibronectin isoform in tumors, Proc. Natl. Acad. Sci. USA {\bf 112}, 8314 (2015).

\item B.N. Mason, A. Starchenko, R.M. Williams, L.J. Bonassar, and C.A. Reinhart-King, Tuning 
three-dimensional collagen matrix stiffness independently of collagen concentration modulates endothelial 
cell behavior, Acta Biomater. {\bf 9}, 4635 (2013).

\item J.C. Kohn, D.W. Zhou, F. Bordeleau, A.L. Zhou, B.N. Mason, M.J. Mitchell, M.R. King, and C.A. 
Reinhart-King, Cooperative effects of matrix stiffness and fluid shear stress on endothelial cell behavior,
Biophys. J. {\bf 108}, 471 (2015).

\item A. Wood, Contact guidance on microfabricated substrata: the response of teleost fin mesenchyme cells 
to repeating topographical patterns, J. Cell Sci. {\bf 90}, 667 (1988).

\item A. Webb, P. Clark, J. Skepper, A. Compston, and A. Wood, Guidance of oligodendrocytes and their 
progenitors by substratum topography, J. Cell Sci. {\bf 108}, 2747 (1995).

\item N. Gomez, S. Chen, and C.E. Schmidt, Polarization of hippocampal neurons with competitive surface
stimuli: contact guidance cues are preferred over chemical ligands, J. Roy. Soc. Interface {\bf 4}, 223 
(2007).

\item A.I. Teixeira, G.A. Abrams, P.J. Bertics, C.J. Murphy, and P.F. Nealey, Epithelial contact guidance 
on well-defined micro- and nanostructured substrates, J. Cell Sci. {\bf 116}, 1881 (2003).

\item W. Loesberg, J. Te Riet, F. van Delft, P. Scon, C. Figdor, S. Speller, J. van Loon, X. Walboomers, 
and J. Jansen, The threshold at which substrate nanogroove dimensions may influence fibroblast alignment and 
adhesion, Biomaterials {\bf 28}, 3944 (2007).

\item N. Nakatsuji and K.E. Johnson, Ectodermal fragments from normal frog gastrulae condition substrata to 
support normal and hybrid mesodermal cell migration in vitro, J. Cell Sci. {\bf 68}, 49 (1984).

\item N. Nakatsuji and K.E. Johnson, Experimental manipulation of a contact guidance system in amphibian 
gastrulation by mechanical tension, Nature {\bf 307}, 453 (1984).

\item A. Ray, Z.M. Slama, R.K. Morford, S.A. Madden, and P.P. Provenzano, Enhanced directional migration of 
cancer stem cells in 3D aligned collagen matrices, Biophys. J. {\bf 112}, 1023 (2017).

\item A. Ray, R. Morford, N. Ghaderi, D. Odde, and P. Provenzano, Dynamics of 3D carcinoma cell invasion into
aligned collagen Integ. Biol. (Camb.) {\bf 10}, 100 (2018).

\item M. Grossman, N. Ben-Chetrit, A. Zhuravlev, R. Afik, E. Bassat, I. Solomonov, Y. Yarden, and I. Sagi, 
Tumor cell invasion can be blocked by modulators of collagen fibril alignment that control assembly of the 
extracellular matrix, Cancer Res. {\bf 76}, 4249 (2016).

\item A. Parekh and A.M. Weaver, Regulation of cancer invasiveness by the physical extracellular matrix 
environment, Cell Adh. Migr. {\bf 3}, 288 (2009).

\item P.P. Provenzano, D.R. Inman, K.W. Eliceiri, S.M. Trier, and P.J. Keely, Contact guidance mediated 
three-dimensional cell migration is regulated by Rho/ROCK-dependent matrix reorganization, Biophys. J. 
{\bf 95}, 5374 (2008).

\item M. Sidani, J. Wycko, C. Xue, J. E. Segall, and J. Condeelis, Probing the microenvironment of mammary 
tumors using multiphoton microscopy, J. Mammary Gland Biol. Neoplasia {\bf 11}, 151 (2006).

\item W. Han {\it et al.}, Oriented collagen fibers direct tumor cell intravasation, Proc. Natl. Acad. Sci. 
USA {\bf 113}, 11208 (2016).

\item D. Truong, J. Puleo, A. Llave, G. Mouneimne, R.D. Kamm, and M. Nikkhah, Breast cancer cell invasion 
into a three dimensional tumor-stroma microenvironment, Sci. Rep. {\bf 6}, 34094 (2016).

\item S.P. Carey, Z.E. Goldblatt, K.E. Martin, B. Romero, R.M. Williams, and C.A. Reinhart-King, Local 
extracellular matrix alignment directs cellular protrusion dynamics and migration through Rac1 and FAK,
Integr. Biol. {\bf 8}, 821 (2016).

\item A.L. Bauer, T.L. Jackson, and Y. Jiang, Topography of extracellular matrix mediates vascular 
morphogenesis and migration speeds in angiogenesis, PLOS Comput. Biol. {\bf 5}, e1000445 (2009).

\item V. Gorshkov, V. Privman, and S. Libert, Lattice percolation approach to 3D modeling of tissue aging,
Physica A {\bf 462}, 207 (2016).

\item M.W. Conklin, J.C. Eickho, K.M. Riching, C.A. Pehlke, K.W. Eliceiri, P.P. Provenzano, A. Friedl, 
and P.J. Keely, Aligned collagen is a prognostic signature for survival in human breast carcinoma. Am. J. 
Pathol. {\bf 178}, 1221 (2011).

\item V. Pavithra, R. Sowmya, S.V. Rao, S. Patil, D. Augustine, V. Haragannavar, and S. Nambiar, 
Tumor-associated collagen signatures: An insight, World J. Dent. {\bf 8}, 224 (2017).

\item A. Brabrand, I.I. Kariuki, M.J. Engstrom, O.A. Haugen, L.A. Dyrnes, B.O. Asvold, M.B. Lilledahl, and 
A.M. Bonfin, Alterations in collagen fibre patterns in breast cancer. A premise for tumour invasiveness?
APMIS {\bf 123}, 1 (2015).

\item J.S. Bredfeldt, Y. Liu, M.W. Conklin, P.J. Keely, T.R. Mackie, and K.W. Eliceiri, Automated 
quantification of aligned collagen for human breast carcinoma prognosis, J. Pathol. Informatics {\bf 5}, 28 
(2014).

\item A.J. Ford and P. Rajagopalan, Extracellular matrix remodeling in 3D: implications in tissue 
homeostasis and disease progression, Nanomed. Nanobiotechnol. {\bf 10}(4), (2018).

\item K.M. Riching {\it et al.}, 3D collagen alignment limits protrusions to enhance breast cancer cell 
persistence, Biophys. J. {\bf 107}, 2546 (2014).

\item P. Friedl and K. Wolf, Tube travel: The role of proteases in individual and collective cancer cell 
invasion, Cancer Res. {\bf 68}, 7247 (2008).

\item E. Schnell, K. Klinkhammer, S. Balzer, G. Brook, D. Klee, P. Dalton, and J. Mey, Guidance of glial 
cell migration and axonal growth on electrospun nanofibers of poly-epsilon-caprolactone and a 
collagen/poly-epsilon-caprolactone blend, Biomaterials {\bf 28}, 3012 (2007).

\item A.A. Saberi, Recent advances in percolation theory and its applications, Phys. Rep. {\bf 578}, 1 
(2015).

\item T.H. Keitt, Spectral representation of neutral landscapes, Landscape Ecol. {\bf 15}, 479 (2000).

\item B.B. Mandelbrot and J.W. van Ness, Fractional Brownian motion, fractional Guassian noise, and their 
applications. SIAM Rev. {\bf 10}, 422 (1968).

\item V. Tejedor, O. \'Benichou, R. Voituriez, R. Jungmann, F. Simmel, C. Selhuber-Unkel, L. B. Oddershede, 
and R. Metzler, Quantitative analysis of single particle trajectories: Mean maximal excursion method, 
Biophys. J. {\bf 98}, 1364 (2010).

\item S. Weber, A.J. Spakowitz, and J.A. Theriot, Bacterial chromosomal loci move subdiffusively through a 
viscoelastic cytoplasm, Phys. Rev. Lett. {\bf 104}, 238102 (2010).

\item G.R. Kneller, K. Baczynski, and M. Pasenkiewicz-Gierula, Consistent picture of lateral subdiffusion 
in lipid bilayers: Molecular dynamics simulation and exact results, J. Chem. Phys. {\bf 135}, 141105 (2011).

\item F. Ghasemi, J. Peinke, M. Sahimi, and M.R. Rahimi Tabar, Regeneration of stochastic processes: An 
inverse method. Europ. Phys. J. B {\bf 47}, 411 (2005).

\item F. Ghasemi, M., Sahimi, J., Peinke, and M.R. Rahimi Tabar, Analysis of non-stationary data for 
heart-rate fluctuations in terms of drift and diffusion coefficients. J. Biol. Phys. {\bf 32}, 117 (2006).

\item H.G. Othmer and A. Stevens, Aggregation, blowup, and collapse: The ABC's of taxis in reinforced random 
walks, SIAM J. Appl. Math. {\bf 57}, 1044 (1997).

\item E.F. Keller and L.A. Segel, Model for chemotaxis, J. Theor. Biol. {\bf 30}, 225 (1971).

\end{list}%

\end{document}